\documentclass[amsmath,amsfonts,amssymb,twocolumn,superscriptaddress,showpacs]{revtex4}

\usepackage{graphicx,psfrag,color}
\usepackage{dcolumn}
\usepackage{bm}
\usepackage{mathbbol}

\newcommand{\br}{{\bf r}}

\begin{document}

\title{Ferrotoroidic Moment as a Quantum Geometric Phase}

\author{C. D. Batista} 
\address{Theoretical Division, Los Alamos National Laboratory, P.O. Box 1663, 
Los Alamos, NM 87545, USA}
\author{G. Ortiz}
\address{Department of Physics, Indiana University, Bloomington,
IN 47405, USA}
\author{A. A. Aligia}
\address{Comisi\'{o}n Nacional de Energ{\'{\i }}a At\'{o}mica, 
Centro At\'omico Bariloche and Instituto Balseiro, \\
8400 S.C. de Bariloche, Argentina}

\date{\today}

\begin{abstract}
We present a geometric characterization of the ferrotoroidic moment
${\boldsymbol \tau}$ in terms of a set of  Abelian Berry phases.   We
also introduce a fundamental complex quantity, $z_{\mu\nu}$,  
which provides an alternative way to calculate ${\boldsymbol \tau}$ and
its moments and is derived from  the tensor $T_{\mu\nu}= 2 \sum_j
r^{\mu}_j S^{\nu}_j$. This geometric framework defines a natural
computational approach for density functional and many-body theories.
\end{abstract}

\maketitle

\textit{Introduction.} The recent discovery of new compounds with strong
multiferroic coupling revised interest in multiferroic materials.  An
attractive component of this discovery is the potential for realizing
strong magnetoelectric effects, i.e., magnetic field (${\bf B}$) induced
electric polarization or electric field (${\bf E}$) induced magnetic
moment. The linear magnetoelectric response is characterized by the
magnetoelectric tensor  ${\hat{\alpha}}$ \cite{Landau}, and different
physical mechanisms can contribute to it. One of these mechanisms has
been recently the focus of considerable attention 
\cite{Schmid01,Fiebig05,Arima05,Sawada05,VanAken07,Ederer07} and is
related to the concept of toroidic moment \cite{Dubovik90}. Materials
with a uniform toroidic moment (ferrotoroidics) exhibit a non-vanishing
magnetoelectric effect, and the ferrotorodic moment is an antisymmetric
component of ${\hat{\alpha}}$. These materials might have technological
applications in the area of computer memories. 

The observation of ferrotoroidic domains in LiCoPO$_{4}$
\cite{VanAken07} confirmed the physical relevance of this concept and
opened the possibility of observing toroidic driven magnetoelectric
effects in certain quantum magnets. However, as it was pointed out
recently \cite{Ederer07}, the toroidic moment is multivalued in a
periodic system. This problem is analogous to the case of the electric
polarization. The root of this indeterminacy lies in the multivalued
nature of the position operator for a periodic system: $\mathbf{r}\equiv
\mathbf{r}+\mbox{\boldmath{$a$}}$, where {\boldmath{$a$}} is a
translation that leaves the system invariant \cite {ours2}. The position
operator appears explicitly in the expressions for the multipolar
moments of charge or current distributions and, thus, all of these
moments are multivalued in a periodic system. In particular, the
toroidic moment appears in a multipolar expansion of a current density
distribution and it couples linearly to $\nabla \times \mathbf{B}$
\cite{Dubovik90}, and to $\mathbf{E}\times \mathbf{B}$ \cite{schmid},
which is easier to control.

Since it is usually convenient to use periodic boundary conditions for
modeling physical systems, it is necessary to know how to compute the
ferrotoroidic moment, ${\boldsymbol\tau }$, and the consequent
magnetoelectric effect within this framework. Although ${\boldsymbol
\tau }$ is multivalued for a periodic system, the change of
${\boldsymbol\tau }$ ($ \Delta {\boldsymbol\tau }$) between two
different physical states is well defined when moving along a particular
path. In the case of the macroscopic electric polarization,
$\mathbf{P}$, of an insulator, the change $\Delta \mathbf{P}$ is
computed by integrating the charge current through a given surface along
an adiabatic path that connects the two different physical states
\cite{ours1,ours2}. This leads to a natural relation between $\Delta
\mathbf{P}$ and the Berry phase associated to an adiabatic evolution  in
an enlarged parameter space \cite{singlep,ours1}, and illustrates the
observable character of the geometric phase. In the same way that the
charge Berry phase $\gamma^{c}$ is a measure of the macroscopic electric
polarization in band or Mott insulators \cite{ours1}, the spin Berry
phase $\gamma^{s}$ can be related to the difference between electrical
polarizations for spin up and down  \cite{gammas}. Here we will
demonstrate that the change in the ferrotoroidic moment, $\Delta
{\boldsymbol\tau }$, can be related to a Berry phase tensor
$\gamma^s_{\mu\nu}$, thus revealing the  geometric character of the
ferrotoroidic response.


\textit{Ferrotoroidic moment.} 
Assume a system of $N$ interacting electrons of mass $m$ and charge $e$
enclosed in a $d$-dimensional box of linear dimension $L_\mu$ and 
volume $\Omega$ ($\mu=1,\cdots,d$). Particle $j$ is labeled by the
coordinates $r^{\mu}_j$ and spin $S^{\nu}_j= \frac{1}{2} \sigma^\nu$
($\sigma^\nu$ are Pauli matrices with $\nu=x,y,z$). The Hamiltonian of
the system is ($\hbar=1$)
\begin{eqnarray}
H^{\lambda} &=& \sum_{i=1}^N
\frac{\mbox{\boldmath{$\Pi$}}_i^2}{2m}+\sum_{i<j} V_{\sf
int}(|\br_i-\br_j|) + \sum_{i=1}^N v_{\sf ext}(\br_i, \lambda(t)) 
\nonumber \\
&+& g_{\sf so} \sum_{i=1}^N 
{\bf U}^{\dagger}(\phi_{\mu\nu}){\boldsymbol \sigma}_i{\bf
U}^{\;}(\phi_{\mu\nu})  \cdot {\bf r}_i \times
\mbox{\boldmath{$\Pi$}}_i,
\label{hamilt}
\end{eqnarray}
where $\mbox{\boldmath{$\Pi$}}_i= {\bf p}_i + {\bf A}^\nu$
is the canonical momentum ($p_\mu=-i \partial_\mu$) with gauge field
${\bf A}^\nu= \frac{\phi_{\mu \nu}}{L_\mu}\sigma^\nu {\hat {\bf
e}}_{\mu}$ ($\hat{\mathbf{e}}_\mu$ is a unit vector along the $\mu$
direction). The flux
$\phi_{\mu\nu}$ is the twist in spin space at the boundary of the box: 
$\Psi_{\alpha}({\bf r}_i+ L_{\mu}{\hat {\bf e}}_{\mu})= \sum_{\beta}
[e^{i\phi_{\mu\nu}\sigma^\nu}]_{\alpha\beta} \Psi_{\beta}({\bf r}_i) $. 

The transformation defined by  ${\bf U}^{\;}(\phi_{\mu\nu})=
e^{i\phi_{\mu\nu}\sigma^\nu \sum_j r_j^{\mu}/L_{\mu}}$ is such that
the eigenstates of $H^\lambda$ satisfy periodic boundary conditions.
The last term in (\ref{hamilt}) corresponds to the relativistic
spin-orbit interaction, while $V_{\sf int}$ and $v_{\sf ext}$ represent
the interaction and  ``external'' potentials, respectively.
The latter depends parametrically on $\lambda(t)$ which is assumed to
change adiabatically in time $t$ from $\lambda(0)=0$ to $\lambda(T)=1$. 

The second quantized form of $H^{\lambda}$ is
$\mathbb{H}^{\lambda}=\int_\Omega d^3 r \ {\cal H}^\lambda(\br)$, with
Hamiltonian density
\begin{eqnarray} 
{\cal H}^{\lambda}(\br) &=&  {\boldsymbol \Psi}^{\dagger}(\br)
\frac{\mbox{\boldmath{$\Pi$}}^2}{2m} {\boldsymbol\Psi}^{\;}(\br) + {\cal
H}_{\sf int}(\br) + {\cal H}^{\lambda}_{\sf ext}(\br)  \nonumber \\
&+& \!\! g_{\sf so}  {\boldsymbol \Psi}^{\dagger}(\br) {\bf
U}^{\dagger}(\phi_{\mu\nu}){\boldsymbol \sigma}{\bf
U}^{\;}(\phi_{\mu\nu}) \cdot {\bf r} \times \mbox{\boldmath{$\Pi$}}
{\boldsymbol \Psi}^{\;}(\br),
\end{eqnarray}
written in terms of the fermionic spinor fields  ${\boldsymbol
\Psi}^{\dagger}(\br)= (\psi^{\dagger}_{\uparrow}(\br) ,
\psi^{\dagger}_{\downarrow}(\br))$,  where
$\psi^{\dagger}_{\alpha}({\br})$ ($\psi^{\;}_{\beta}({\br})$)  creates
(annihilate) an electron with spin $\alpha (\beta)=\{\uparrow,\downarrow
\}$ at position $\br$. 

The spin contribution to the ferrotoroidic moment $\tau_\nu$ is defined
as  the antisymmetric component of the tensor $t_{\mu\nu}$. The
$t_{\mu\nu}$-density  field is defined by the local condition $\sum_\mu
\partial_\mu{\cal T}_{\mu\nu}(\br)= - 2 {\cal S}_{\nu}(\br)$. For a
finite system with open boundary conditions (surfaces), the following
relation holds as long as ${\cal T}_{\mu\nu}(\br)$ vanishes outside the
system
\begin{equation}
t_{\mu \nu} = \frac{T_{\mu \nu}}{\Omega}= \frac{2}{\Omega}  \int_\Omega 
d^3r \ r_{\mu} \ {\cal S}_{\nu}({\br})=\frac{1}{\Omega}  \int_\Omega 
d^3r \ {\cal T}_{\mu\nu}({\br}) ,
\label{tmunu}
\end{equation}
written in terms of the spin density field
\begin{equation} 
{\cal S}_{\nu}({\br})=\sum_{\alpha \beta} S^{\nu}_{\alpha \beta}({\br})
, \  {\cal S}^{\nu}_{\alpha \beta}({\br}) = \frac{1}{2} 
\psi^{\dagger}_{\alpha}({\br}) \sigma^{\nu}_{\alpha \beta}
\psi^{\;}_{\beta}({\br}),
\end{equation}
where $\int_\Omega  d^3r  \ {\cal S}_{\nu}({\bf r})=0$ to have $t_{\mu
\nu}$ independent of the origin of coordinates \cite{note}. More
specifically  ${\tau_\eta} = \frac{\mu_B}{4}\sum_{\mu\nu}
\epsilon_{\eta\mu\nu} t_{\mu \nu}$, with $\epsilon_{\eta\mu\nu}$ the
Levi-Civita tensor. Like for the electric polarization
\cite{ours1,ours11}, the third member of Eq. (\ref{tmunu}) is not well
defined for a system with periodic boundary conditions because the
position operator is not well defined
\cite{ours2}. However, the last member of Eq. (\ref{tmunu}) is still
well defined,  and starting from such expression for $t_{\mu\nu}$, we
will demonstrate that the change $\langle \Delta t_{\mu \nu} \rangle$
between two different physical states  can be obtained from a set of
Abelian Berry phases. 

Since the ground state (GS) of $H^\lambda$ evolves continuously between
$\lambda(0)=0$ and $\lambda(T)=1$, the change is given by
\begin{eqnarray}
\langle \Delta {t_{\mu \nu}} \rangle &=& \frac{1}{\Omega}  \int_0^{1} d\lambda
\int_{\Omega} d^3r   \ \partial_{\lambda} \langle \Phi^{\lambda}_0 |  {\cal
T}_{\mu \nu}({\bf r})  |\Phi^{\lambda}_0 \rangle,
\label{DTmunu}
\end{eqnarray}
where  $|\Phi^{\lambda}_0 \rangle$ is the GS of $\mathbb{H}^{\lambda}$
with flux $\phi_{\mu\nu}$, and  $\langle \Delta {t_{\mu \nu}} \rangle=
\langle\Phi^{\lambda(T)}_0 |t_{\mu \nu}| \Phi^{\lambda(T)}_0\rangle  -
\langle\Phi^{\lambda(0)}_0 |t_{\mu \nu}| \Phi^{\lambda(0)}_0\rangle$. 
By using adiabatic perturbation theory \cite{ours1,thouless} up to first
order in the time derivative we obtain
\begin{eqnarray}
&& \!\!\!\!\!\!\!\!\!\!\!\!\! \frac{1}{\Omega}  \int_{\Omega} d^3r   \
\partial_{\lambda} \langle \Phi^{\lambda}_0 |  {\cal T}_{\mu \nu}({\bf r}) 
|\Phi^{\lambda}_0\rangle = \nonumber \\
&& \!\!\!\!\!\!\!\! \!\!\!\!\! -\frac{iL_{\mu}}{\Omega}\!\! 
\sum_{m\neq0}   \frac{\langle \Phi^{\lambda}_0|\partial_{\phi_{\mu
\nu}}{\mathbb H}^{\lambda}| \Phi^{\lambda}_m\rangle  \langle
\Phi^{\lambda}_m|\partial_{\lambda}{\mathbb H}^{\lambda}|
\Phi^{\lambda}_0\rangle} {[E^{\lambda}_0-E^{\lambda}_m]^2} + {\rm c.
c.},
\label{DTmunu2}
\end{eqnarray}
where we have used that ${\mathbb H}^{\lambda}|\Phi^{\lambda}_m\rangle=
E^{\lambda}_m|\Phi^{\lambda}_m\rangle$ ($\langle \Phi^{\lambda}_m |
\Phi^{\lambda}_{m'}\rangle = \delta_{m,m'}$) and
\begin{equation}
i [{\mathbb H}^{\lambda}, {\cal T}_{\mu \nu} ({\bf r}) ] =
\partial_t {\cal T}_{\mu \nu}=  
L_{\mu} \partial_{\phi_{\mu \nu}} {\cal H}^{\lambda}({\bf r}),
\label{e2}
\end{equation}
In the absence of the spin-orbit term,  $\partial_t {\cal T}_{\mu \nu}=2
{\cal J}^s_{\mu \nu}({\bf r})$ with ${\cal J}^s_{\mu \nu}({\bf r})$ the
density of spin current \cite{note2} satisfying the continuity equation
$\sum_{\mu} \partial_{\mu} {\cal J}^s_{\mu \nu}({\bf r}) +  \partial_t
{\cal S}_{\nu}({\bf r})=0$. By using the relations   $\langle
\Phi^{\lambda}_m|\partial_{\lambda}{\mathbb H}^{\lambda}| 
\Phi^{\lambda}_0\rangle = \langle
\Phi^{\lambda}_m|[\partial_{\lambda},{\mathbb H}^{\lambda}]|
\Phi^{\lambda}_0\rangle$, and $ \langle
\Phi^{\lambda}_0|\partial_{\phi_{\mu\nu}}{\mathbb H}^{\lambda}|
\Phi^{\lambda}_m\rangle = \langle
\Phi^{\lambda}_0|[\partial_{\phi_{\mu\nu}},{\mathbb H}^{\lambda}]|
\Phi^{\lambda}_m\rangle$,  we obtain from Eqs. (\ref{DTmunu}) and
(\ref{DTmunu2})
\begin{eqnarray}
\langle \Delta {t_{\mu \nu}} \rangle \!\! &=& \!\!
-\frac{i L_{\mu}}{\Omega}  \int_0^{1} \!\! d\lambda \sum_{m\neq 0} \langle
\partial_{\phi_{\mu \nu}}\Phi^{\lambda}_0| \Phi^{\lambda}_m\rangle 
\langle \Phi^{\lambda}_m|\partial_{\lambda} \Phi^{\lambda}_0\rangle
+{\rm c.c} \nonumber \\
&=&\frac{L_{\mu}}{\Omega} \int_{0}^{1}  d\lambda \ {\cal
B}({\bf \xi}) , \ \ \ {\xi}=(\phi_{\mu\nu},\lambda) ,
\label{DTmunu3}
\end{eqnarray}
with ${\cal B}({\xi})=i(\langle \partial_{\lambda} \Phi^{\lambda}_0 |
\partial_{\phi_{\mu\nu}} \Phi^{\lambda}_0 \rangle \! - \! \langle
\partial_{\phi_{\mu \nu}}\Phi^{\lambda}_0|\partial_{\lambda}
\Phi^{\lambda}_0\rangle )$. Since $t_{\mu \nu}$ is a bulk property (and
we assumed that there is no level crossing for the GS as a function of
$\phi_{\mu \nu}$), its value should not depend on the boundary
conditions when we take the thermodynamic limit. Thus \cite{note}
\begin{eqnarray}
\langle \Delta {t_{\mu \nu}} \rangle &=&\!\! \frac{L_{\mu}}{2 \pi
\Omega} \int_{0}^{1} \!\!\! d\lambda  \int_0^{2\pi} \!\!\!\!\!\!
d\phi_{\mu\nu} \ {\cal B}({\xi})=\frac{L_{\mu}}{2 \pi
\Omega}\oint_{\Gamma} {\cal A} ({\xi}) \cdot d {\xi} \nonumber 
\end{eqnarray}
with the line integral performed along the contour $\Gamma$ of 
$[0,2\pi]\times$[0,1] in the plane $(\phi_{\mu\nu},\lambda)$, and  ${\bf
\cal A}(\xi)=i  \langle \Phi_0^\lambda | \nabla_{\xi} \Phi_0^\lambda
\rangle$. Assuming a path-independent gauge \cite{ours1}
\begin{eqnarray}
\langle \Delta {t_{\mu \nu}} \rangle &=&  \frac{L_{\mu}}{2\pi\Omega} \
[\gamma^s_{\mu\nu}(1) - \gamma^s_{\mu\nu}(0)] \nonumber \\
\mbox{with } \gamma^s_{\mu\nu}(\lambda) &=& i\int_0^{2\pi} \!\!\!
d\phi_{\mu\nu} \ \langle \Phi^{\lambda}_0| \partial_{\phi_{\mu \nu}}
\Phi^{\lambda}_0\rangle . 
\label{gmunu}
\end{eqnarray}
One can also introduce fluxes along space directions other than $\mu$
and, as long as the gap does not close, average over these additional
fluxes. The Berry phases  $\gamma^s_{\mu\nu}$ are anholonomies
associated with the parallel transport of a vector state (GS) in the
parameter space determined by the fluxes. Equation (\ref{gmunu}) shows
that the {\it quantum of uncertainty}, related to the lack of history in
the adiabatic evolution, is at least the inverse of the total transverse
section $L_{\mu}/{\Omega}$. In a periodic system with primitive unit
cell volume $\Omega_0 < \Omega$, the quantum is  larger and equal to
$L^0_{\mu}/\Omega_0$, where $L^0_{\mu}$ is the length of the unit cell
along the $\mu$-direction \cite{ours1}.  

The extension to lattice systems is straightforward
\begin{eqnarray}
\mathbb{H}^L= \! \! \! \!\sum_{{\bf r} {\bf r'},\alpha\beta} \!\! t_{{\bf
r}{\bf r'}}  \left (c_{{\bf r}\alpha}^{\dagger}  \left[ e^{i \theta_{\mu
\nu}^{{\bf r}{\bf r'}}{\sigma }^{\nu}} \right]_{\alpha
\beta} c_{\mathbf{\bf r'}\beta }^{\;}+\mathrm{H.c.} \right )
+ \mathbb{H}^L_{\rm int},
\label{hub}
\end{eqnarray}
where $c_{\br\alpha}^{\dagger}$ creates an electron of spin $\alpha$ at
lattice site $\br$, $\theta_{\mu \nu}^{{\bf r}{\bf r'}}= \int_{\bf
r}^{\bf r'} dr_{\mu} A_{\mu \nu}= \phi_{\mu \nu}
(r'_{\mu}-r_{\mu})/L_{\mu}$,  $n_{\bf r}=\sum_{\alpha}c_{{\bf
r}\alpha}^{\dagger } c_{{\bf r}\alpha }^{\;}$.  The spin current on the
bond  $({\bf r},{\bf r} +\hat{\mathbf{e}}_{\mu })$ is  $ {\cal
J}^{s}_{\mu \nu}=\frac{i \hat{\mathbf{e}}_{\mu }}{2}\sum_{\alpha \beta}
t_{{\bf r}{\bf r}+\hat{\mathbf{e}}_{\mu }}  (c_{{\bf r}
+\hat{\mathbf{e}}_{\mu }\alpha }^{\dagger }c_{{\bf r}\beta }^{\;}-c_{
{\bf r}\alpha }^{\dagger }c_{\mathbf{{\bf r}+\hat{e}_{\mu }}\beta
}^{\;})\sigma _{\alpha \beta }^{\nu}$. The charge Berry phase
$\gamma^c$  is obtained when ${\theta_{\mu \nu }^{{\bf r}{\bf
r'}}{\sigma}^{\nu }}$ is replaced  by $\int_{\bf r}^{\bf
r'}dr_{\mu} A_{\mu}  = \phi_{\mu}^c (r'_{\mu}-r_{\mu})/L_{\mu}$, i.e., 
the flux associated with the usual vector potential $A_{\mu}$ of the
electromagnetic field. 

{\it Localization Indicators}. There is a formal mathematical connection
between the geometric phases $(\gamma^c,\gamma^s)$ and the localization
indicators $(z_{L}^{c},z_{L}^{s})$ \cite{ours2,ours3,ours4} that were
introduced to discriminate between different phases such as conductors
and insulators \cite{res}. $z_{L}^{c}$ is defined by Eq. (\ref{zmunu})
below,  replacing  ${\varphi _{\mu \nu }{\boldsymbol\sigma }^{\nu }}$
with a scalar coupling ${\varphi}^c_{\mu}$.  For a  one dimensional
system along the $x$ direction and spin quantization axis $z$,
$z_{L}^{s}$ corresponds to the component $z_{x z}$ of the tensor defined
by Eq. (\ref{zmunu}).  The phase of the  localization number $z_{L}^{k}$
(with $k=c,s$), is related to the Berry phase $\gamma^{k}$ through the
relation $\gamma^{k}={\rm{Im}}\ln z_{L}^{k}$. 

Different indicators provide complementary information. They  may
display different convergence properties to the thermodynamic limit, 
and most importantly, some indicators are more akin to Monte Carlo
methods while others are geared towards numerical renormalization group
approaches. The localization parameters $z_{L}^{k}$ and their related
Berry phases $\gamma^{k}$ have also been used as {\it sharp} topological
indicators
for establishing quantum phase diagrams of interacting systems
\cite{ours11,ours4,hir,ab,abn,nak}.

By analogy to $z_{L}^{c,s}$, define the localization indicators
\begin{equation}
z_{\mu \nu }[\varphi _{\mu \nu }]=\langle \Psi _{0}|\Psi _{0}(\varphi
_{\mu \nu })\rangle \ ,\ |\Psi _{0}(\varphi _{\mu \nu })\rangle =e^{i
\varphi _{\mu \nu} T_{\mu \nu }}|\Psi _{0}(\mathbf{0})\rangle \ ,  
\label{zmunu}
\end{equation}
where $\varphi _{\mu \nu }={2\pi }/L_{\mu}$  and $|\Psi_{0}\rangle $ is
the GS \cite{note}. 

Considered as a continuous function of $\varphi_{\mu \nu }$,  $z_{\mu
\nu}$  plays the role of a characteristic function generating all
moments of the $T_{\mu \nu }$ tensor, and trivially satisfies $z_{\mu
\nu }[0]=1$, $|z_{\mu \nu }[\varphi _{\mu \nu }]|\leq 1$, $z_{\mu \nu
}[\varphi _{\mu \nu }]=z_{\mu \nu }^{\ast }[-\varphi _{\mu \nu }]$.
Note, however, that the operator $T_{\mu \nu }$ is not a genuine
operator in the Hilbert space bundle defined above, although its
exponential is a legitimate one. Therefore, expectation values of
arbitrary powers of $T_{\mu \nu }$ have only meaning in terms of $z_{\mu
\nu }$ (i.e., there is a \textit{quantum of uncertainty} \cite{ours1}).
Assuming analyticity in the neighborhood of $ \varphi _{\mu \nu }=0$ 
(i.e., the system has a gap in the thermodynamic
limit), $z_{\mu \nu }$ can be written in terms of cumulants $
C_{k}(T_{\mu \nu })$ 
\begin{equation}
z_{\mu \nu }=\exp \left[ \sum_{k=1}^{\infty }\frac{(i\varphi _{\mu \nu
})^{k}}{ k!}C_{k}(T_{\mu \nu })\right] \ .
\end{equation}
with the end result that 
\begin{eqnarray}
\varphi _{\mu \nu }^{-1}\ \mathrm{Im}\ln z_{\mu \nu } &=&\langle T_{\mu
\nu }\rangle +\mathcal{O}(\varphi _{\mu \nu }^{2})\ ,  \label{tdelz} \\
-\varphi _{\mu \nu }^{-2}\ \ln |z_{\mu \nu }|^{2} &=&\langle T_{\mu \nu
}^{2}\rangle -\langle T_{\mu \nu }\rangle ^{2}+\mathcal{O}(\varphi _{\mu
\nu }^{2})\ .
\end{eqnarray}
so that the phase of $z_{\mu \nu}$ is related to the tensor $T_{\mu
\nu}$ while its modulus provides information on its quantum 
fluctuations  (it vanishes in the thermodynamic limit for systems with
gapless excitations that have a non-zero spin current).   

To illustrate an elementary application of Eq. (\ref{tdelz}),  assume
that $|\Psi_{0}\rangle $ is the GS of a periodic system with the
primitive unit cell of volume $\Omega_0$, (such as that of LiCoPO$_{4}$ 
\cite{VanAken07} but with spins pointing in the $z$ direction)
consisting of two spin up and two down in the direction $z$, and
displaced in the direction $x$ so that  the contribution to $T_{xz}$
from each unit cell,  $\tilde{t}_{xz}=2 \sum_{j=1}^{4} x_j S^{z}_j \neq
0$ ($\langle t_{\mu \nu } \rangle =  \tilde{t}_{\mu\nu}/\Omega_0$).
Thus,  the first member of Eq. (\ref{tdelz}) should give $N_u
\tilde{t}_{\mu \nu }$ where  $N_u$ is the number of unit cells. This
equation is invariant if any coordinate $r_{j}^{\mu }$ is replaced by
$r_{j}^{\mu }+nL_{\mu }$ with $n$ integer. Equation (\ref{zmunu}) gives
$z_{\mu \nu }=\exp (2\pi i N_u \tilde{t}_{\mu \nu }/L_{\mu })$, and
replacing into Eq. (\ref{tdelz}), $\langle T_{\mu \nu }\rangle =N_u
\tilde{t}_{\mu \nu }$ as expected. Note that the same result is obtained
if $\tilde{t}_{\mu \nu }$ is replaced by $\tilde{t}_{\mu \nu }+nq$,
where $n$ is integer and $q=L_{\mu }/N_u$ is the quantum of uncertainty
in $\tilde{t}_{\mu \nu }$. However, for translational invariant systems,
one can calculate $z_{\mu\nu}$ in the effective one-dimensional problem
with fixed total transversal wave vector, ($K_y$, $K_z$), for which
$q=L^0_{\mu}$.


\textit{Effective single-body schemes.} One can implement previous ideas
in the framework of spin density (matrix) functional or Hartree-Fock
theories. The Hamiltonian acting on the Kohn-Sham (or Hartree-Fock)
orbitals, $\psi^\alpha_{n\mathbf{k}}(\mathbf{r})$, with lattice
periodicity {\boldmath{$a$}} 
\begin{eqnarray}
\sum_\beta (\frac{\mathbf{p}^2}{2m}
\delta_{\alpha\beta}+ \hat{v}_{\alpha\beta}
(\mathbf{r}))
\psi^\beta_{n\mathbf{k}}(\mathbf{r})
=\epsilon_{n\mathbf{k}}\psi^\alpha_{n\mathbf{k}}(\mathbf{r}) 
\label{KS}
\end{eqnarray}
defines the two-component spinor $\Psi_{n\mathbf{k}}(\mathbf{r})=\binom{
\psi^\uparrow_{n\mathbf{k}}(\mathbf{r})}{\psi^\downarrow_{n\mathbf{k}} (
\mathbf{r})}$ with the following generalized Bloch-spinor conditions 
\begin{eqnarray}
{\boldsymbol \Psi}_{n\mathbf{k}}(\mathbf{r} + a_\mu
\hat{\mathbf{e}}_\mu)&=& e^{i (k_{\mu} a_{\mu} +
\frac{\phi_{\mu\nu}}{L_{\mu}} a_\mu {\sigma}^\nu)} 
{\boldsymbol \Psi}_{n\mathbf{k}}(\mathbf{r})  \nonumber \\
{\boldsymbol \Psi}_{n\mathbf{k}}(\mathbf{r})&=& e^{i (\mathbf{k}+ 
\frac{\phi_{\mu\nu}}{L_{\mu}}{\sigma}^\nu
\hat{\mathbf{e}}_{\mu})  \cdot \mathbf{r}}  {\boldsymbol
u}_{n\mathbf{k}}(\mathbf{r})
\end{eqnarray}
where ${\boldsymbol u}_{n \mathbf{k}}(\mathbf{r} +
\mbox{\boldmath{$a$}})={\boldsymbol u}_{n\mathbf{k }}(\mathbf{r})$ is
the periodic part of the Bloch-spinor. In Eq. (\ref{KS})
$\hat{v}_{\alpha\beta}$ represents the effective one-body potential
(which includes the external, Hartree, exchange and correlation
components). To simplify the final expression of $\langle t_{\mu
\nu}\rangle$, it is convenient to express the spinor ${\boldsymbol
u}_{n\mathbf{k}}$ in the  basis of eigenstates of ${
\sigma}^{\nu}$, 
\begin{eqnarray}
{\boldsymbol u}_{n\mathbf{k}}=\begin{pmatrix} {u}_{n\mathbf{k}}^{\nu +}
\cr {u}_{n\mathbf{k}}^{\nu -} \end{pmatrix} , \mbox{ with } {
\sigma}^{\nu} {\boldsymbol u}_{n\mathbf{k}}=\begin{pmatrix} 
+{u}_{n\mathbf{k}}^{\nu +} \cr -{u}_{n\mathbf{k}}^{\nu -} 
\end{pmatrix}, 
\end{eqnarray}
and the periodic part of the (non-interacting) many-particle
wavefunction  along each of the spin directions is the product of Slater
determinants: $\Phi^\nu_{\mathbf{k}}({\bf r}_1,\cdots,{\bf r}_{N})={\sf
Det}(A^{\nu +}({\bf k})) {\sf Det}(A^{\nu -}({\bf k}))$, where  $A^{\nu
\theta}_{ij}({\bf k})={u}^{\nu \theta}_{i \mathbf{k}}(\mathbf{r}_j)$
with $\theta=\pm$, $1 \leq i \leq N_{\theta}$,  and  $1\leq j \leq
N_{+}$ ($1+N_{+}\leq j \leq N$) for $\theta=+$ ($\theta=-$).
$N=N_+ + N_-$ is the number of particles in the unit cell.  Then, using
Eq. (\ref{gmunu})

\begin{eqnarray}
\langle t_{\mu\nu}(\lambda) \rangle= \frac{i}{(2 \pi)^3} \ \sum_{\theta=\pm}
\sum_{n=1}^{\mathrm{ N}_{\theta}} \int_{\mathrm{BZ}} d^3k \
\theta \langle {u}^{\nu \theta}_{n \mathbf{k}} | \ \partial_{k_{\mu}}
{u}^{\nu \theta}_{n  \mathbf{k}} \rangle  ,
\end{eqnarray}
with the momentum integral evaluated over the Brillouin zone (BZ)
corresponding to the periodicity {\boldmath${a}$}. As mentioned above,
what has physical meaning is the {\it change} in the tensor
$t_{\mu\nu}$. Therefore, $\Delta t_{\mu\nu} =
t_{\mu\nu}(1)-t_{\mu\nu}(0)$. Here, as in the case of charge and spin
macroscopic polarizations \cite{ours1}, the price paid for considering a
two-point formula (i.e., forgetting about the $\lambda$-dependent path)
is the appearance of a quantum of uncertainty. 

\textit{Geometry of magnetoelectric response.} Finally, to understand
the geometric content of the magnetoelectric response, we need to
understand the Riemannian structure of our Hilbert space bundle.
Consider a set of normalized states $\{|\Psi_{0}(
\mbox{\boldmath{$\varphi$}} ) \rangle \}$, where 
$\mbox{\boldmath{$\varphi$}}$ represents $3d$ real numbers $\varphi_{\mu
\nu }$. Let's assume that this manifold of quantum states is generated
by the action of the group of transformations $|\Psi_{0}(
\mbox{\boldmath{$\varphi$}} )\rangle =e^{i\mbox{\boldmath{$\varphi$}}
\cdot \mathbf{T}}|\Psi _{0}(0)\rangle$, where
$\mbox{\boldmath{$\varphi$}}\cdot \mathbf{T}=\sum_{\mu\nu}\varphi _{\mu
\nu } T_{\mu \nu }$. The expectation value of the generalized 
\textit{twist} operator $e^{i\mbox{\boldmath{$\varphi$}}\cdot
\mathbf{T}}$  measures the character of the spectrum of low-energy
$T_{\mu \nu}$-excitations. Can we find a measure of the
\textit{distance} between two of these quantum states?

Provost and Vallee addressed the problem of establishing a Riemannian
structure on an arbitrary differentiable manifold of quantum states
\cite{metric}. 
Berry, later on, extended their work
by introducing a geometric tensor \cite{Berry5} whose physical relevance
in the context of electric charge polarization was addressed in Refs.
\cite{ours3,martin} and also for spin polarization in Ref. \cite{ours3}.
We now introduce a
general quantum geometric tensor 
\begin{eqnarray}
G_{\mu \nu; \alpha\beta}[\mbox{\boldmath{$\varphi$}}]&=& \langle
\partial_{\varphi_{\mu\nu}} \Psi_0 | \hat{P}_0[\mbox{\boldmath{$\varphi$}}] |
\partial_{\varphi_{\alpha\beta}}\Psi_0 \rangle ,
\end{eqnarray}
whose real and imaginary parts $G_{\mu \nu; \alpha\beta}[\mbox{\boldmath{$
\varphi$}}]=g_{\mu \nu; \alpha\beta}[\mbox{\boldmath{$\varphi$}}] + i \
\Omega_{\mu \nu; \alpha\beta}[\mbox{\boldmath{$\varphi$}}]$ can be written 
\begin{eqnarray}
\left \{ 
\begin{array}{l}
g_{\mu \nu; \alpha\beta}[\mbox{\boldmath{$\varphi$}}]= \mathrm{Re}
\langle \partial_{\varphi_{\mu\nu}}\Psi_0|
\partial_{\varphi_{\alpha\beta}} \Psi_0 \rangle -
\gamma_{\mu\nu}[\mbox{\boldmath{$\varphi$}}] \gamma_{\alpha\beta}[
\mbox{\boldmath{$\varphi$}}] \ , \  \\ 
\\ 
\Omega_{\mu \nu;\alpha\beta}[\mbox{\boldmath{$\varphi$}}]= \mathrm{Im}
\langle \partial_{\varphi_{\mu\nu}} \Psi_0 |
\partial_{\varphi_{\alpha\beta}} \Psi_0 \rangle ,
\end{array}
\right .
\end{eqnarray}
with $ \hat{P}_0[\mbox{\boldmath{$\varphi$}}]= {\mathbb{1}} - |\Psi_0(
\mbox{\boldmath{$\varphi$}}) \rangle \langle
\Psi_0(\mbox{\boldmath{$\varphi$}}) |$, and
$\gamma_{\mu\nu}[\mbox{\boldmath{$\varphi$}}] = i \langle \Psi_0 |
\partial_{\varphi_{\mu\nu}} \Psi_0 \rangle$ the Berry connection
\cite{ours1}. The real part of $G_{\mu \nu;
\alpha\beta}[\mbox{\boldmath{$\varphi$}}]$ is a symmetric and positive
definite tensor representing a generalization of the metric introduced
in Ref. \cite{metric}. Moreover, it is interesting to remark that the
infinitesimal distance is related to the quantum fluctuations of the
$T_{\mu\nu}$ tensor, i.e., $g_{\mu
\nu,\alpha\beta}(\mbox{\boldmath{$\varphi$}}\rightarrow\mathbf{0}) =
\langle T_{\mu\nu} T_{\alpha\beta}\rangle - \langle T_{\mu\nu} \rangle
\langle T_{\alpha\beta} \rangle $, with expectation values evaluated
over $| \Psi_0(\mathbf{0}) \rangle \equiv | \Psi_0 \rangle$. In a sense,
the metric structure on the manifold is fixed by the quantum
fluctuations which determine the modulus of $ z_{\mu\nu}$ in the
thermodynamic limit. On the other hand, the antisymmetric tensor
$\Omega_{\mu \nu;\alpha\beta}[\mbox{\boldmath{$\varphi$}}] = \mathrm{Im}
\langle \partial_{\varphi_{\mu\nu}} \Psi_0 |
\partial_{\varphi_{\alpha\beta}} \Psi_0 \rangle$ ($\Omega_{\mu
\nu;\alpha\beta}[\mbox{\boldmath{$\varphi$}}]=-
\Omega_{\alpha\beta;\mu\nu}[\mbox{\boldmath{$\varphi$}}]$) plays the
role of a curvature, and is a quantity connected to the non-dissipative
part of the spin conductance in adiabatic transport.

In summary, we have introduced a formalism that not only leads to  a
geometric  understanding of a spin component of the magnetoelectric
response but  also provides a computational method to study changes in
ferrotoroidic moment, $\Delta{\boldsymbol \tau}$, when the system under
consideration has periodic boundary conditions (no surface). Forgetting
about the history of the evolution leading to $\Delta{\boldsymbol \tau}$
amounts to  the appearance of a quantum of uncertainty of magnitude
$\frac{\mu_B}{4} \frac{L^0_\mu}{\Omega_0}$.  Our many-body
formalism generalizes, giving an operational physical interpretation to,
the concept of spin Berry phase introduced in  \cite{gammas}, and should
be particularly useful  within the framework of spin density (matrix)
functional or Hartree-Fock theories.  

LANL is supported by US DOE under Contract No. W-7405-ENG-36.

\end{document}